\documentclass[reprint,secnumarabic,amssymb, aps, prl,superscriptaddress]{revtex4}
\usepackage{graphicx}
\usepackage{amsfonts,amsmath,amssymb}

\begin{document}
\title{Room temperature chiral magnetic skyrmion in  ultrathin magnetic nanostructures }


\author{Olivier Boulle}
\affiliation{Univ. Grenoble Alpes, SPINTEC, F-38000 Grenoble, France}
\affiliation{CNRS, SPINTEC, F-38000 Grenoble, France}
\affiliation{CEA, INAC-SPINTEC, F-38000 Grenoble, France}
\author{Jan Vogel}
\affiliation{CNRS, Institut N\'{e}el,  25 avenue des Martyrs, B.P. 166, 38042 Grenoble Cedex 9, France}
\affiliation{Univ. Grenoble Alpes, Institut N\'{e}el,  25 avenue des Martyrs, B.P. 166, 38042 Grenoble Cedex 9, France}
\author{Hongxin Yang}
\affiliation{Univ. Grenoble Alpes, SPINTEC, F-38000 Grenoble, France}
\affiliation{CNRS, SPINTEC, F-38000 Grenoble, France}
\affiliation{CEA, INAC-SPINTEC, F-38000 Grenoble, France}
\author{Stefania Pizzini}
\affiliation{CNRS, Institut N\'{e}el,  25 avenue des Martyrs, B.P. 166, 38042 Grenoble Cedex 9, France}
\affiliation{Univ. Grenoble Alpes, Institut N\'{e}el,  25 avenue des Martyrs, B.P. 166, 38042 Grenoble Cedex 9, France}
\author{Dayane de Souza Chaves}
\affiliation{CNRS, Institut N\'{e}el,  25 avenue des Martyrs, B.P. 166, 38042 Grenoble Cedex 9, France}
\affiliation{Univ. Grenoble Alpes, Institut N\'{e}el,  25 avenue des Martyrs, B.P. 166, 38042 Grenoble Cedex 9, France}
\author{Andrea Locatelli}
\affiliation{Elettra-Sincrotrone, S.C.p.A, S.S 14 - km 163.5 in AREA Science Park 34149
Basovizza, Trieste, Italy}
\author{Tevfik Onur Mente\c{s}}
\affiliation{Elettra-Sincrotrone, S.C.p.A, S.S 14 - km 163.5 in AREA Science Park 34149
Basovizza, Trieste, Italy}
\author{Alessandro Sala}
\affiliation{Elettra-Sincrotrone, S.C.p.A, S.S 14 - km 163.5 in AREA Science Park 34149
Basovizza, Trieste, Italy}
\author{Liliana D. Buda-Prejbeanu}
\affiliation{Univ. Grenoble Alpes, SPINTEC, F-38000 Grenoble, France}
\affiliation{CNRS, SPINTEC, F-38000 Grenoble, France}
\affiliation{CEA, INAC-SPINTEC, F-38000 Grenoble, France}
\author{Olivier Klein}
\affiliation{Univ. Grenoble Alpes, SPINTEC, F-38000 Grenoble, France}
\affiliation{CNRS, SPINTEC, F-38000 Grenoble, France}
\affiliation{CEA, INAC-SPINTEC, F-38000 Grenoble, France}
\author{Mohamed Belmeguenai}
\affiliation{LSPM (CNRS-UPR 3407), Universit\'{e} Paris 13, Sorbonne Paris Cit\'{e}, 99 avenue Jean-Baptiste Cl\'{e}ment, 93430 Villetaneuse, France}
\author{Yves Roussign\'{e}}
\affiliation{LSPM (CNRS-UPR 3407), Universit\'{e} Paris 13, Sorbonne Paris Cit\'{e}, 99 avenue Jean-Baptiste Cl\'{e}ment, 93430 Villetaneuse, France}
\author{Andrey Stashkevich}
\affiliation{LSPM (CNRS-UPR 3407), Universit\'{e} Paris 13, Sorbonne Paris Cit\'{e}, 99 avenue Jean-Baptiste Cl\'{e}ment, 93430 Villetaneuse, France}
\author{Salim Mourad Ch\'{e}rif}
\affiliation{LSPM (CNRS-UPR 3407), Universit\'{e} Paris 13, Sorbonne Paris Cit\'{e}, 99 avenue Jean-Baptiste Cl\'{e}ment, 93430 Villetaneuse, France}
\author{Lucia Aballe}
\affiliation{ALBA Synchrotron Light Facility, Carretera BP 1413, Km. 3.3, Cerdanyola del Vallès, Barcelona 08290, Spain}
\author{Michael Foerster}
\affiliation{ALBA Synchrotron Light Facility, Carretera BP 1413, Km. 3.3, Cerdanyola del Vallès, Barcelona 08290, Spain}
\author{Ma\"{i}rbek Chshiev}
\affiliation{Univ. Grenoble Alpes, SPINTEC, F-38000 Grenoble, France}
\affiliation{CNRS, SPINTEC, F-38000 Grenoble, France}
\affiliation{CEA, INAC-SPINTEC, F-38000 Grenoble, France}
\author{St\'{e}phane Auffret}
\affiliation{Univ. Grenoble Alpes, SPINTEC, F-38000 Grenoble, France}
\affiliation{CNRS, SPINTEC, F-38000 Grenoble, France}
\affiliation{CEA, INAC-SPINTEC, F-38000 Grenoble, France}
\author{Ioan Mihai Miron}
\affiliation{Univ. Grenoble Alpes, SPINTEC, F-38000 Grenoble, France}
\affiliation{CNRS, SPINTEC, F-38000 Grenoble, France}
\affiliation{CEA, INAC-SPINTEC, F-38000 Grenoble, France}
\author{Gilles Gaudin}
\affiliation{Univ. Grenoble Alpes, SPINTEC, F-38000 Grenoble, France}
\affiliation{CNRS, SPINTEC, F-38000 Grenoble, France}
\affiliation{CEA, INAC-SPINTEC, F-38000 Grenoble, France}

\begin{abstract}

Magnetic skyrmions are chiral spin structures with a whirling configuration. Their topological properties, nanometer size and the fact that they can be moved by   small current densities  have opened   a new paradigm for the manipulation of magnetisation at the nanoscale. To date,    chiral skyrmion structures   have been experimentally demonstrated only  in   bulk materials and in epitaxial ultrathin films  and under  external magnetic field or  at low temperature. Here, we report on the observation of stable skyrmions in sputtered ultrathin Pt/Co/MgO nanostructures, at room temperature and zero applied  magnetic field. We use high lateral resolution  X-ray magnetic circular dichroism  microscopy  to image  their   chiral N\'{e}el internal structure which we explain as due to the large strength of the Dzyaloshinskii-Moriya interaction as revealed by spin wave spectroscopy measurements.  Our results are substantiated by   micromagnetic simulations and numerical models, which allow  the identification of the physical mechanisms governing  the size and stability of the skyrmions.
\end{abstract}

\maketitle
\

The recent discovery of nanometer size whirling magnetic structures  named magnetic skyrmions  has opened a new path to manipulate magnetisation at the nanoscale~\cite{Bogdanov89,Roesler06,Muehlbauer09,Pappas09,Yu10,Nagaosa13,Fert13}. A key feature of such magnetic nano-objects is their  chiral and topologically non-trivial spin structure, i.e   their magnetisation texture cannot be continuously transformed into the uniform magnetic state without causing a singularity~\cite{Braun12AP}. For a surface $C$, this topological property is characterized by the skyrmion number $S=\frac{1}{4\pi}\int_C \mathbf{m}.(\frac{\partial \mathbf{m}}{\partial x}\times\frac{\partial \mathbf{m}}{\partial y})dxdy $ ($m$ is the unit magnetization vector). For a single  skyrmion, $|S|$ is quantized and equal to $1$   whereas $S=0$ for the ferromagnetic state or any topologically trivial structure.
The skyrmion chiral property is generally driven~\cite{Roesler06,Nagaosa13} by  an additional anti-symmetric term in the exchange energy, namely the Dzyaloshinskii-Moriya interaction (DMI), induced by the lack of structural inversion symmetry and the spin orbit coupling (SOC)~\cite{Moriya60,Dzyaloshinskii57SPJ}. This  additional energy term writes  $E_{DM}=\mathbf{D}.(\mathbf{S_1}\times\mathbf{S_2})$ where    $\mathbf{D}$ is the DMI characteristic vector and $\mathbf{S_1}$ and  $\mathbf{S_2}$ are two neighbouring spins. Thus, the DMI   tends to make the magnetisation rotate around  $\mathbf{D}$.  An additional important feature is that skyrmions can be manipulated by in-plane currents~\cite{Jonietz10,Schulz12,Yu12,Jiang15} which has led to novel concepts of non-volatile magnetic memories~\cite{Fert13,Koshibae15} and logic devices~\cite{Zhang15c} where skyrmions in nanotracks are the information carriers.  The  nm size of the skyrmion  combined with the low current density needed to induce its motion~\cite{Jonietz10,Yu12,Jiang15} would  lead to devices with unprecedented combination of high storage density,  large data transfer rate and low power consumption~\cite{Fert13,Koshibae15}.  
The topological ($|S|=1$) and chiral properties of   skyrmions    are  keys  for such devices as they control the  current  induced skyrmion dynamics~\cite{Moutafis09,Buettner15,Fert13,Sampaio13,Tomasello14,Jiang15} and the interaction of skyrmions with other spin structures and   the  nanostructure edges~\cite{Iwasaki13,Sampaio13,Tomasello14,Zhou14b, Zhang15c,Zhang15f}.

Although  predicted   at the end of the 80's~\cite{Bogdanov89}, chiral skyrmion lattices and isolated skyrmions were  observed   only recently    in   B-20-type   bulk  or thin film  chiral  magnets~\cite{Nagaosa13} such as    MnSi~\cite{Muehlbauer09}, Fe$_{1-x}$Co$_x$Si~\cite{Yu10b}, FeGe~\cite{Yu11}, Mn$_{1-x}$Fe$_x$Ge~\cite{Shibata13} and in ultrathin magnetic films composed of Fe or PdFe monolayers on Ir(111)~\cite{Heinze11,Romming13,Romming15}. 
In B-20 magnets,  the  DMI results from the    non-centrosymmetric crystalline structure and Bloch-like skyrmions   have been observed   below room temperature~\cite{Yu11} and in the presence of an external magnetic field~\cite{Nagaosa13}. 
In epitaxial  heavy metal/ferromagnetic ultrathin films~\cite{Heinze11,Romming13,Romming15}, the DMI arises from the breaking of inversion symmetry at the interfaces  combined with the large spin-orbit coupling in the heavy metal~\cite{Fert13}, leading to  skyrmions with a N\'{e}el-like structure~\cite{Bogdanov94,Bogdanov99}. However, the observed skyrmions were stable only at low temperature~\cite{Sonntag14} which prevents any use for applications. In addition, the ultrathin films  were grown by molecular beam epitaxy which is not suitable for industrial production.

Recently, the attention has shifted to   ultrathin ferromagnetic/heavy metal  films  deposited by sputtering~\cite{Monso02}. This class of materials combines several features which makes their use  appealing for the study of skyrmion structures and their applications: i/ The  magnetic parameters controlling the skyrmion stability and size, i.e the anisotropy, the DMI and the exchange~\cite{Bogdanov94,Rohart13}, can be easily tuned by playing with the nature and thickness of the materials composing the multilayers. ii/ They are characterized by  a large DMI~\cite{ Freimuth14b,Emori14a, Pizzini14,Di15,Nembach15NP,Stashkevich15PRB,Di15,Belmeguenai15,Yang15b} which leads  to chiral N\'{e}el domain walls (DWs)~\cite{Chen13,Tetienne15}. iii/ Large   current induced spin orbit torques are present~\cite{Miron11,Garello13}  which results in fast current induced DW motion~\cite{Miron11a,Thiaville12}. iv/ The deposition by sputtering is fast and spatially homogeneous  and is compatible with standard spintronics devices such as magnetic tunnel junctions, which makes the industrial integration  straightforward. Whereas several recent experimental works have studied magnetic bubbles in such materials~\cite{Buettner15,Moreau-Luchaire15,Woo15,Jiang15}, and demonstrated their current induced motion~\cite{Woo15,Jiang15}, the direct evidence of their chiral internal structure     is still lacking. Here  we report on the observation of   stable chiral skyrmions   in   sputtered ultrathin  Pt/Co(1~nm)/MgO nanostructures  at room temperature and zero applied magnetic field. We used photoemission electron microscopy combined with X-ray magnetic circular dichroism (XMCD-PEEM)    to demonstrate their  chiral N\'{e}el  internal structure. The XMCD-PEEM   combines several advantages   for the observation of magnetic nanostructures, such as skyrmions: firstly,  a high lateral spatial resolution (down to 25~nm); secondly,  the magnetic contrast is  proportional to the projection of the local magnetisation along the X-ray beam direction. In our experiment, the X-ray beam   impinges at a grazing angle of 16 degrees on the sample surface plane   so that the contrast is approximately three times larger for the in-plane component of the magnetisation than for the out-of-plane one. This important feature allows the direct imaging of  the internal in-plane spin structure of    DWs or skyrmions.

\begin{figure}[!h]
	\centering
		\includegraphics[width=0.75\textwidth]{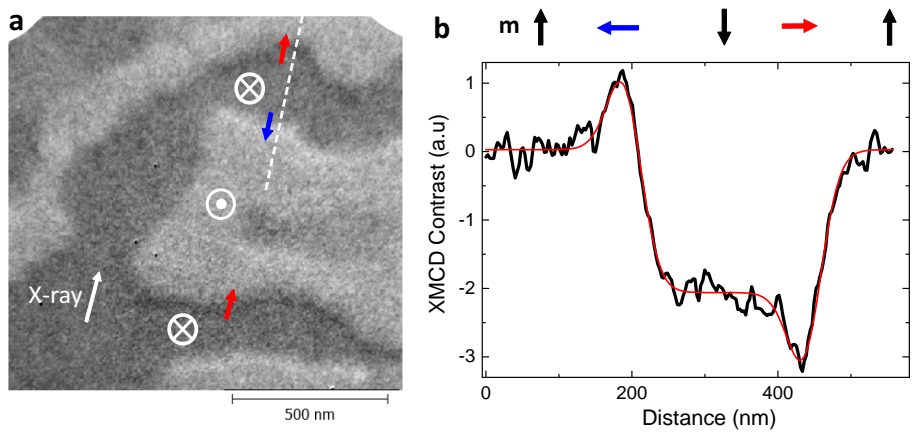}
	\caption{\textbf{Imaging of the chiral N\'{e}el  structure of domain walls  using XMCD-PEEM magnetic microscopy.} (a) Magnetic image of a multidomain state in the continuous Pt/Co/MgO film.  For DWs lying perpendicular  to the X-ray beam direction, thin white and black lines can be seen, corresponding to the magnetisation being aligned antiparallel and parallel to the photon beam respectively. This demonstrates  their chiral N\'{e}el structure. (b)  Linescan of the magnetic contrast corresponding to the dotted white line in (a). To reduce the noise, the contrast has been averaged perpendicularly to the linescan over  60 nm.  The red line is   a fit assuming a chiral N\'{e}el DW structure convoluted by a Gaussian function to take into account the finite spatial resolution~\cite{SuppMat}.} 
	\label{Fig1}
\end{figure}

\section*{Observation of chiral N\'eel domain walls using XMCD-PEEM}
All images shown  here were acquired at room temperature   and, unless otherwise stated, no  external  magnetic field  was  applied during the experiments. The PEEM observations were done in a virgin demagnetized state obtained   after  the sample nanofabrication and annealing. Complementary magnetisation measurements on unpatterned thin films show that the sample is   magnetised perpendicularly to the film in the magnetic domains, which is due to a large interfacial uniaxial    anisotropy.    Figure~\ref{Fig1}(a) shows an XMCD-PEEM magnetic image   of a multidomain state in the continuous film. Dark and bright grey regions correspond  respectively to the magnetisation pointing down and up. Interestingly, we observe a sharp increase in the dichroic contrast for DWs perpendicular to the X-ray beam, with a strong dark contrast when going from an down  to a up magnetized domain (along the beam direction) and a strong bright contrast when going from a up to a down magnetized domain.  This can be seen more easily in the linescan of the magnetic contrast shown in Fig.~\ref{Fig1}(b), corresponding to the white dashed line in Fig.~\ref{Fig1}(a). A peak in the contrast is observed at the up/down DW position while a dip is observed at the down/up DW position. Thus, the magnetisation in the up/down DW is aligned antiparallel to the in-plane direction of the X-ray beam whereas the magnetisation in the down/up DW is aligned parallel. We conclude that the DW magnetisation  is perpendicular to the DW surface with an opposite magnetisation direction for the two DWs.  This demonstrates that DWs in this material are chiral N\'{e}el DWs with a left-handed chirality. Note that for Bloch DWs, the magnetisation would be  always perpendicular to the beam direction so that no peak or dip in the magnetic contrast should be observed. The linescan is  well fitted  assuming a chiral N\'{e}el DW structure, the finite resolution of the instrument   being modeled by a Gaussian convolution (red curve, Fig.~\ref{Fig1}(b)). The fit leads to a  DW width of  $29.5\pm4$~nm ($\pi\sqrt{A/K_{eff}}$)~\cite{SuppMat}.   

\begin{figure}[!h]
	\centering
		\includegraphics[width=0.75\textwidth]{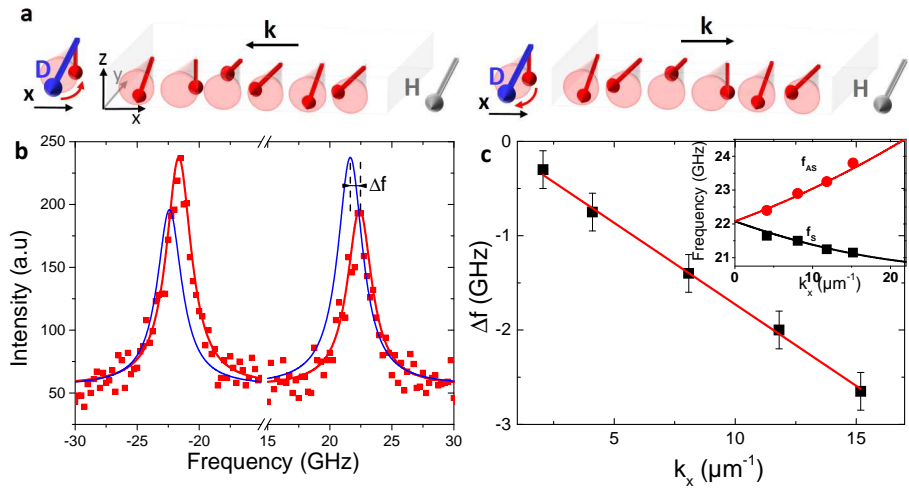}
	\caption{\textbf{Brillouin Light Scattering spinwave spectroscopy.} (a) Principle of the measurement. At a given time $t$, when moving along x, the magnetisation rotates  clockwise(/counterclockwise) around the D vector for spin waves propagating along x (-x), which leads to different DM energy.   (b) BLS spectra for an in-plane magnetic field H=0.7~T and $k_{x}=4.1 \mu$m$^{-1}$. The red squares  are  experimental data whereas the red lines are Lorentzian fits. The  blue line is a Lorentzian fit of the experimental data  inverted with respect to $f=0$. (c)  $\Delta f=f_S-f_{AS}$ as a function $k_{x}$ for $H=0.7$~T, where  $f_{S}$ and $f_{AS}$ are respectively the Stokes and Anti-Stokes resonance frequency. Inset: $f_{S}$ and $f_{AS}$  as a function $k_x$.}
	\label{Fig2}
\end{figure}

\section*{Spin wave spectroscopy experiments and \emph{ab-initio} calculations}

The driving force of the DW and skyrmion chiral structure is the DMI. To further quantify its amplitude in our films, we measured   the frequency shift of oppositely propagating spin  waves using  spin wave spectroscopy experiments~\cite{Cortes-Ortuno13,Moon13, Di15,Nembach15NP,Stashkevich15PRB,Belmeguenai15}.   The idea of the measurement is the following:  When the magnetisation is pulled in the plane by an  external  magnetic field $H_y$, the $D$ vector is oriented  along y for spin waves propagating along the x axis (see Fig.~\ref{Fig2}(a)). Thus, at a given time $t$, when moving along the x axis,  the magnetisation rotates anticlockwise around the $D$ vector for    spin waves with  $k_x<0$ and clockwise for $k_x>0$. This leads to an energy shift   for spin waves with opposite $k_x$ vector due to the DMI and  the  corresponding frequency shift writes $\Delta f(k_x)=f(k_x)-f(-k_x)=2\gamma k_x D/(\pi M_s)$. To measure $\Delta f$, we have carried out spin wave spectroscopy using the Brillouin Light Scattering technique in a backscattering   geometry~\cite{Belmeguenai15}.  A spin wave spectrum is shown on Fig.~\ref{Fig2}(b) (red dots) for an in-plane magnetic field of 0.7~T and  $k_{x}=4.1 \mu$m$^{-1}$. The  Stokes (S) and Anti - Stokes(AS)   peaks are observed, corresponding to $\pm k_x$.  The blue line is a Lorentzian fit of the experimental data  inverted with respect to $f=0$,   which shows that the Stokes peak has a   frequency  different to the Anti-Stokes peak,  as is expected in the presence of DMI. The shift in frequency $\Delta f=f_S-f_{AS}$ scales linearly with  $k_{x}$  (Fig.~\ref{Fig2}(c)), which allows to extract a DM parameter $D=2.05\pm0.3$~mJ/m$^2$. Note that the sign of $\Delta f$    is consistent with the sign of $D$ and the left handedness of the N\'{e}el DW we observe experimentally.   As $D$ is expected to be inversely proportional to the film thickness $t$~\cite{Belmeguenai15}, one can define  a related interfacial DM parameter $D_s$ such that $D=D_s/t$ and we find a value $D_s=2.17\pm0.14$~pJ/m. To our knowledge, this  value is  the highest reported  so far for a  sputtered magnetic ultrathin film.  To better understand this large value, we carried out \emph{ab-initio} calculations of the DMI in  Pt/Co(n ML)/vacuum and Pt/Co(n ML)/MgO multilayers~\cite{SuppMat,Yang15b}. For n=5 ML of Co,  equivalent to a total Co thickness of 1~nm, the \emph{ab-initio} calculations predict  $D=2.3$~mJ/m$^2$ in relatively good agreement with experiments. Note that a   lower value $D=1.5$~mJ/m$^2$ is predicted for a Pt/Co[N]/vacuum structure, which underlines the role played by the Co/oxide interface and in particular the large electric field due to the charge transfer between O and Co atoms.

\section*{Room-temperature skyrmion in a  magnetic nanostructure}

\begin{figure}[!h]
	\centering
		\includegraphics[width=0.75\textwidth]{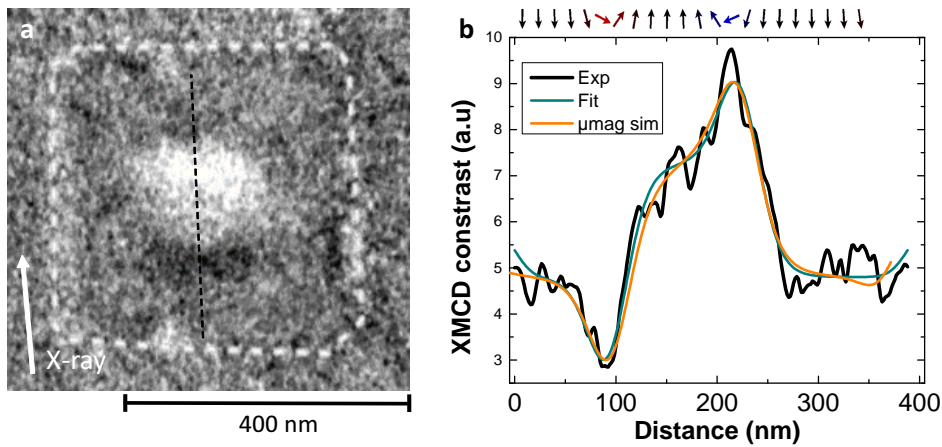}
	\caption{\textbf{Magnetic skyrmion observed at room temperature and zero applied external magnetic field} (a) XMCD-PEEM  image of  a 420~nm square dot (indicated by the dotted line) and (b) linescan along the dotted black line  (black line). The linescan has been averaged perpendicularly to the linescan over 30 nm. The blue line is a fit to the data using a Gaussian convoluted $360^\circ$ DW profile~\cite{SuppMat}.   The orange line is the contrast predicted by the micromagnetic simulations. }
	\label{Fig3}
\end{figure}

In continuous thin films, isolated magnetic bubbles~\cite{Slonczewski79} or chiral skyrmions~\cite{Nagaosa13} have   been observed so far in the presence of a perpendicular magnetic field, which  breaks the stripe domains or helical structure driven by the magnetostatic  or DM energy. However, it is known that   single magnetic bubbles can be stabilized without external magnetic fields using geometrical confinement in patterned nanostructures~\cite{Hehn96S, Moutafis07PRB}. Here we patterned different structures with various sizes and shapes (circular, square) in our Pt/Co/MgO thin films. Fig.~\ref{Fig3}(a) shows a circular magnetic domain stabilized in the middle  of a 420 nm wide square dot,    imaged at room temperature and no applied magnetic field. As   observed in the multidomain structure,  a sharp black/white contrast    is observed  at the DW position at the bottom/top of the  central domain. This leads to a dip/peak in the dichroic contrast   when doing a linescan   along the   domain diameter in the beam direction  (see Fig.~\ref{Fig3}(b)). This indicates that the in-plane DW magnetisation is aligned parallel/antiparallel to the X-ray beam at the bottom/top of the reversed domain, i.e the DW surrounding the circular domain is a chiral N\'{e}el DW. This chiral border leads to a skyrmion number   $|S|=1$  for this structure. This demonstrates that the observed circular domain is  a N\'{e}el like magnetic skyrmion.  To extract the size of the skyrmion from the image, we assume that the magnetisation   profile   can be described by a $360^\circ$ DW profile~\cite{Braun94,Kubetzka03,Romming15} : $\mathbf{m}=\sin\theta(r)\mathbf{u_r}+\cos\theta(r)\mathbf{u_z}$ with $\theta(r)=\theta_{DW}(r-d/2)+\theta_{DW}(r+d/2)$, where $\theta_{DW}(r)=2\arctan[\exp(r/\Delta)]$;  $d$ is the skyrmion diameter and $\Delta$ is the DW width. The unit vector $\mathbf{u_r}$ is the polar unit vector.
The blue curve (Fig.~\ref{Fig3}(b)) shows a fit of the experimental linescan assuming  a Gaussian convoluted $360^\circ$ DW and a good agreement is obtained with experimental data. From the fit, a skyrmion diameter $d=130\pm2.5$~nm is extracted. 

\begin{figure}[!h]
	\centering
		\includegraphics[width=0.6\textwidth]{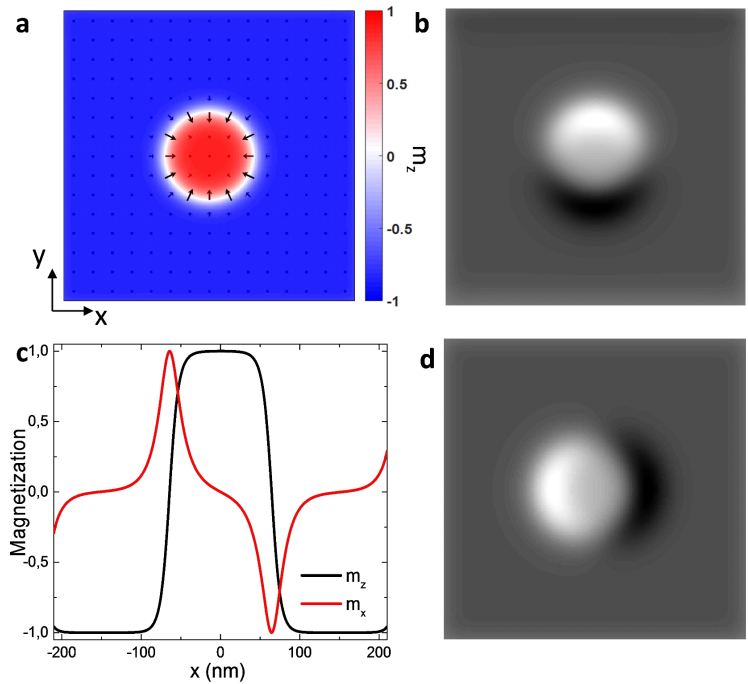}
	\caption{\textbf{ Micromagnetic simulations.} (a) Magnetisation distribution and (b) simulated magnetic contrast  of the magnetic bubble in a 420~nm dot. (c) Magnetisation component $m_x$ and  $m_z$ along the dotted white line in (a). (d) XMCD-PEEM   contrast assuming a Bloch bubble structure. For the simulated magnetic contrast image,  a spatial lateral resolution of 28~nm is assumed  as  obtained from the fit of the topological image of Fig.~\ref{Fig3}~\cite{SuppMat}.}
	\label{Fig4}
\end{figure}

To better understand these experimental results, we carried out micromagnetic simulations using  experimental values for the   magnetocrystalline  anisotropy constant $K$,   the magnetic moment per surface area and the DMI interaction amplitude $D$. The exchange constant was used as a free parameter  and the best agreement between experiments and micromagnetic simulations is obtained for $A=27.5$~pJ/m, a value in line with previous measurements of $A$ in ultrathin magnetic multilayers~\cite{Metaxas07}. Using these parameters, micromagnetic simulations predict a stable left-handed  chiral skyrmion structure at zero external magnetic field (see Figure~\ref{Fig4}(a)) with a diameter of  128~nm and a DW width of 37~nm. From this magnetisation pattern, an experimental magnetic image can be simulated and a good agreement is obtained with the experimental results (see Fig.~\ref{Fig4}(b)  and Fig.~\ref{Fig3}(b), orange curve for the simulated linescan).  We show  on Fig.~\ref{Fig4}(d) the same reconstructed experimental image assuming a Bloch DW structure. The image is rotated $90^\circ$ with respect to the chiral N\'{e}el bubble structure and it is in clear disagreement with our experimental data. Finally, the simulations allow  us to reconstruct the structure of the observed skyrmion. We show in Fig.~\ref{Fig4}(c) a linescan of the in-plane ($m_x$) and out-of-plane ($m_z$) component of the magnetisation along the skyrmion diameter, as predicted by the micromagnetic simulation. The skyrmion diameter ($\sim130$~nm) being large compared to    the DW width (37~nm), the magnetisation profile is close to two independent chiral N\'{e}el DWs, as can be seen on the $m_z$ profile.  
 
\section*{Numerical calculations}

\begin{figure}[!h]
	\centering
		\includegraphics[width=0.6\textwidth]{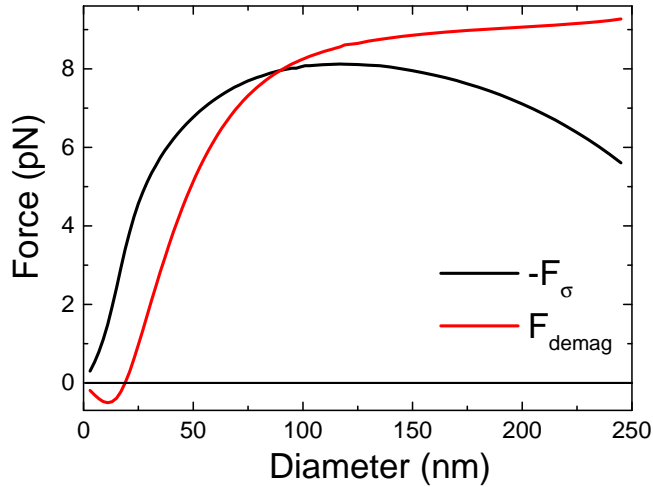}
	\caption{\textbf{Forces acting on the DW as a function of the skyrmion diameter.} $F_\sigma$ is the force due to the exchange and effective anisotropy (black curve), $F_{mag}$ is the force due to the magnetostatic interaction between the domains (red curve).  }
	\label{Fig5}
\end{figure}

The observation of   stable skyrmions at zero external magnetic field   raises the question of the physical mechanisms that govern   the skyrmion stability and size in our experiments.    To address this point,    we consider a simple model where the magnetisation  in the dot $\theta(r)$  is described by a circular $360^\circ$  N\'{e}el  DW profile. The free energy  $E$ in the circular dot-shaped nanostructure  can be written as the sum of two terms~\cite{Bogdanov89,Bogdanov94,Kiselev11,Rohart13}: (1) the skyrmion energy $E_{\sigma}[\theta(r)]$ due to the exchange, anisotropy   and internal DW stray field energies and (2)  the  energy due to the magnetostatic interactions between the domains $E_{mag}$. Assuming a radial symmetry, $E_{\sigma}[\theta(r)]$ can be written as~\cite{Bogdanov89,Bogdanov94,Kiselev11,Rohart13}:  
\begin{equation}
E_{\sigma}[\theta(r)]=2\pi t\int_0^R\left\{A\left[\left(\frac{d\theta}{dr} \right)^2+\frac{\sin^2\theta}{r^2} \right]-D\left[\frac{d\theta}{dr}+\frac{\cos\theta\sin\theta}{r} \right]+ (K_{eff} + E^{s}_{DW})\sin^2\theta\right\}rdr 
\end{equation}

where $K_{eff}=K-\mu_0M_s^2(1-N_{DW})/2$ is the effective anisotropy~\cite{Braun94,Braun99JAP} ($K$ is  the magneto-crystalline  anisotropy constant), $t$ is the film thickness, $R$ the dot radius.  The   demagnetizing energy due to the magnetic charges within the DWs $E^{s}_{DW}$  is described by a constant demagnetizing factor $N_{DW}$ such that $E^{s}_{DW}=N_{DW}\mu_0M_s^2/2$.   The energies $E_{\sigma}$ and $E_{mag}$~\cite{Guslienko15} can be evaluated  as a function of the skyrmion diameter $d$ assuming a  $360^\circ$ DW profile and a 420 nm diameter circular dot  (the magnetic   parameters correspond  to the experiment  of Fig~\ref{Fig3} , see methods).  More physical insight is  obtained from the effective forces  $F_\sigma(d)=-\frac{\partial E_{\sigma}}{\partial d}$ and $F_{mag}(d)=-\frac{\partial E_{mag}}{\partial d}$ which are plotted in Fig.~\ref{Fig5}. A first interesting feature is that $F_\sigma(d)$ cancels out for $d\sim20$~nm, which thus would   be an equilibrium size for the skyrmion in the absence of the domain magnetostatic energy. This equilibrium is the result of a balance between the DW energy cost which is proportional to $d$ and tends to decrease the skyrmion diameter and the curvature energy cost due the 
exchange energy which scales as $1/d$~\cite{Rohart13}. 
 However, the magnetostatic force $F_{mag}$ is large enough at low diameter to destabilize this balance and the final equilibrium position is obtained for a   larger value  of $d\sim90$~nm, where the two forces are equal. This underlines that the magnetostatic energy plays an important role in the stability and size of the skyrmion at zero external magnetic field. We also carried out micromagnetic simulations for square nanostructures with larger lateral dimensions. We observed that for sides larger than 1.2~$\mu$m,  the skyrmion structure is not stable and a stripe domains structure appears~\cite{SuppMat}. This may explain why we did not observe any skyrmions but stripe domains  for larger structures with  1~$\mu$m sides~\cite{SuppMat}. The confinement is thus an additional important feature for the skyrmion stability.

It may seem surprising that the domain magnetostastic energy plays  such an important role in the skyrmion stability given  the very small thickness of the layer. Indeed, magnetostatic effects are expected to favor multidomain states  only when the film thickness $t$ is of the order or larger than the characteristic length $l_w=\sigma/(\mu_0M_s^2)$, where $\sigma$ is the DW surface  energy. This criterion expresses the competition between the cost in the DW energy and the gain in the magnetostatic energy when creating a domain.   The    energy $\sigma$ scales as    $\sigma\sim4\sqrt{AK_{eff}}-\pi D$  so that it is decreased in the presence of  DMI.   Thus, the criterion  $l_w\sim t$ can be fulfilled even for very low thickness.  In our experiments, we find that $l_w$  is indeed of the order of the film thickness, $l_w\sim1$~nm.  This leads to the counterintuitive conclusion that  the higher the DMI, the larger  the role  played by the  magnetostatic interaction in the skyrmion stability. Note that this analysis holds for $\sigma>0$, i.e   $D<D_c$ where $D_c=4\sqrt{AK_{eff}}/\pi$, which is the case in our sample where $D/D_c=0.8$.

To conclude, we have observed  stable   magnetic skyrmions at room temperature in   ultrathin Pt/Co/MgO films  in the absence of applied external  magnetic field. Using the high resolution XMCD-PEEM magnetic microscopy technique, we were able to demonstrate its internal left-handed chiral N\'{e}el structure, which can be explained by the large Dzyaloshinskii-Moriya interaction measured in this material. Micromagnetic   simulations are in agreement with our experimental results and we show that the equilibrium skyrmion structure is the result  of a balance between the DW energy modulated by the DMI and the magnetostatic interaction in the patterned structure. This balance is   very sensitive to the different magnetic parameters that are $D$, $K$, $A$ and the total magnetic moment. The     lateral confinement   also plays an important role since it   leads to  a modulation of the  magnetostatic energy. Hence, the size and stability of magnetic skyrmions can be finely tuned by playing with these different parameters. This  will be a key for the design of future devices based on the manipulation of magnetic skyrmions.

\section*{Methods}
\textbf{Sample preparation and magnetic microscopy experiment}
The Ta(3)/Pt(3)/Co(0.5-1)/MgO$_x$/Ta(1) (thickness in nm) film was deposited by magnetron sputtering on a 100~mm high resistivity Si wafer, then annealed for 1.5~h at $250^\circ$~C under vacuum and under an in-plane magnetic field of $\mu_0H=240$~mT.   The Co layer was deposited as a wedge using a rotating cover during the deposition. The nominal thickness at the position of observation is $t=0.98$~nm. The samples were patterned in different shapes (circle, square) and sizes   using standard nanofabrication techniques. The XMCD-PEEM magnetic microscopy experiments were  carried out with the SPELEEM III microscope (Elmitec GmbH) at the Nanospectroscopy beamline~\cite{Mentes14BJN} at the Elettra synchrotron in Basovizza, Trieste, Italy and at the CIRCE beamline with the collaboration of ALBA staff~\cite{Aballe15JoSR}. To fit the linescan of Fig.~\ref{Fig1}(b), the  standard deviation of the spatial Gaussian convolution $\sigma$ is used as a free parameter~\cite{SuppMat} leading to $2\sigma=40$~nm~. In Fig~\ref{Fig3}(b), $\sigma$ is deduced from a fit with an error function of a   linescan  of the topological image of the dot , which leads to $2\sigma=28$~nm.

\textbf{Brillouin light scattering experiment}
The Brillouin light scattering experiments setup and conditions  are the same as described in Ref.~\cite{Belmeguenai15}. A  backscattering geometry has been used. The investigated spin wave vector lies in the plane of incidence and its length is $k_x = 4\pi \sin(\theta_{inc})/\lambda$ (with $\theta_{inc}$ the angle of incidence and $\lambda=532$ nm the wavelength of the illuminating laser). The external magnetic field was applied perpendicular to the incidence plane, which allows spin waves propagating along the in-plane direction perpendicular to the applied field to be probed (Damon-Eshbach geometry).

\textbf{Micromagnetic simulations}
The micromagnetic simulations were carried out using different micromagnetic codes: a homemade code ~\cite{Buda_02CMS}, the Mumax3 code~\cite{Vansteenkiste11}, the OOMMF code~\cite{Donahue99}. The following parameters were used~\cite{SuppMat} : $K=1.45\times10^6$ J/m$^3$, $M_s=1.4\times10^6$~A/m, $A=27.5$ pJ/m, $D=2.05$~mJ/m$^2$ and a film thickness $t=1.06$~nm. The lateral size of  the elementary cells was typically between 1 and 3~nm. The  results presented in this paper  were obtained for a lateral cell size of 1~nm. In Fig.~\ref{Fig4}(b), a spatial Gaussian convolution with standard deviation $\sigma=14$~nm was used to simulate the finite lateral spatial resolution of the microscope.

\textbf{Numerical calculations of the skyrmion energy}
 $E_{\sigma}(d)$ was calculated numerically by minimizing $E_{\sigma}$  with respect to $\Delta$ at fixed $d$. $E_{mag}(d)$  was then evaluated from the total magnetostatic energy $E_{mag0}$ of a Bloch DW in a dot  using Ref.~\cite{Guslienko15}.
\begin{eqnarray*} 
E_{mag}(d)=E_{mag0}-  t(1- N_{DW})\frac{\mu_0M_s^2}{2}\int_0^R\cos^2\theta\enspace2\pi r dr \enspace \mathrm{with } \\
	E_{mag0}=\frac{4\pi t}{R}\int_0^\infty  [1-\exp(-\beta x)  ]I^2(x) dx\\
	I(x)=\int_0^1dr' r' J_0(xr') \cos\theta(r')
\end{eqnarray*}

where $r'=r/R$ and $J_0$ is the Bessel function of order zero. $R$ is the dot radius.
$N_{DW}=0.0188$ was deduced from the  width of the DW predicted by the micromagnetic simulations.

 \paragraph{Competing Interests}
 The authors declare that they have no competing financial interests.

\paragraph{Acknowledgements} The authors would like to thank  Andr\'{e} Thiaville, Murat Cubukcu, Lorenzo Camosi, Michael Caminale and   W. Savero-Torres for discussions and their help in experiments. For their contribution to the CIRCE beamline at the Alba synchrotron, we  would like to thank  C. Escudero, V. Perez-Dieste, E. Pellegrin, J. Nicolas and S. Ferrer. S.P. and J.V. acknowledge the support of the Agence Nationale de la Recherche, project ANR-14-CE26-0012 (ULTRASKY).
\paragraph{Author Contributions} O.B.  conceived and designed the experiments. O.B., J.V., S.P., D.S.C., A.L, T.O.M., A.S., L.A., M.F. participated in the XMCD-PEEM experiments. O.B and J.V. analyzed the microscopy data. H.Y. and M.C. carried out the \emph{ab-initio} calculations. O.B., L.B-D. and O.K. carried out the micromagnetic simulations, O.B. carried out the numerical calculations, S.A. deposited the magnetic multilayers, M.B., Y.R., A.S. carried out the BLS experiments. O.B. wrote the manuscript. All authors discussed the results and commented on the manuscript.
\paragraph{Correspondence} Correspondence and requests for materials should be addressed to O.B. \newline  (email: olivier.boulle@cea.fr).




\end{document}